\documentclass[11pt]{article}
\usepackage{amssymb}
\usepackage{amsfonts}
\usepackage{amsmath}
\usepackage{hyperref}

\makeatletter \@addtoreset{equation}{section} \makeatother

\newtheorem{theorem}{Theorem}
\newtheorem{lemma}{Lemma}

\newtheorem{proposition}{Proposition}

\begin{document}

\title{Central limit theorem for linear eigenvalue statistics of the Wigner and sample covariance random matrices}
\author{ M. Shcherbina
\\
Institute for Low Temperature Physics, Ukr. Ac. Sci \\
 47 Lenin ave, 61135 Kharkov Ukraine, e-mail: Shcherbi@ilt.kharkov.ua}
\date{}

\maketitle

\begin{abstract} We consider two classical ensembles of the random matrix theory: the Wigner matrices and
 sample covariance matrices, and prove Central Limit Theorem for linear eigenvalue statistics under  rather weak
(comparing with results known before) conditions  on the number of derivatives of the test functions and also on the
number of the entries moments. Moreover, we develop a universal method which allows one to obtain automatically  the
bounds for the variance of differentiable test functions, if there is a bound for the variance of the trace of the
resolvent of random matrix. The method is applicable not only to the Wigner and sample covariance matrices, but to any
ensemble of random matrices.
\end{abstract}

%

\section{Introduction}\label{s:1}
The Wigner Ensembles for real symmetric matrices  is  a family of $n\times n$ real symmetric matrices $M$ of the form
\begin{equation}
M=n^{-1/2}W,  \label{MW}
\end{equation}%
where $W=\big\{w_{jk}^{(n)}\big\}_{j,k=1}^{n}$ with $w_{jk}^{(n)}=w_{kj}^{(n)}\in \mathbb{R}$, $1\leq j\leq k\leq n$,
and  $w_{jk}^{(n)}$, $1\leq j\leq k\leq n$ are independent random variables such that
\begin{equation}
\mathbf{E}\big\{w_{jk}^{(n)}\big\}=0,\quad \mathbf{E}\big\{(w_{jk}^{(n)})^{2}\big\}=1, \quad j\not=k,\quad
\mathbf{E}\big\{(w_{jj}^{(n)})^{2}\big\}=w_2.
  \label{Wmom12}
\end{equation}%
Here and below we denote  $\mathbf{E}\{.\}$ the averaging with respect to all random parameters of the problem. Let
$\{\lambda_j^{(n)}\}_{i=1}^n$ -- be eigenvalues of $M$. Since the pioneer work of Wigner \cite{W} it is known that if
we consider the linear eigenvalue statistic corresponding to any continuous test function $\varphi$:
\begin{equation}\label{Linst}
\mathcal{N}_n[\varphi]=\sum_{j=1}^n\varphi(\lambda_j^{(n)}),
\end{equation}
 then $n^{-1}\mathcal{N}_n[\varphi]$ converges in probability to the limit
\begin{equation}\label{conv_sc}
\lim_{n\to\infty}n^{-1}\mathcal{N}_n[\varphi]=\int\varphi(\lambda)\rho_{sc}(\lambda)d\lambda.
\end{equation}
where $\rho_{sc}(\lambda)$ is the famous semicircle density
\[\rho_{sc}(\lambda)=\frac{1}{2\pi}\sqrt{4-\lambda^2}\mathbf{1}_{[-2,2]}.\]
The result of this type, which is the analog of the Law of Large Numbers  of the classical probability theory, normally
is the first step in studies of the eigenvalue distribution for any ensemble of random matrices. For the Wigner
ensemble
 this result, obtained initially in \cite{W} for Gaussian $W=\big\{w_{jk}^{(n)}\big\}_{j,k=1}^{n}$, was improved in
\cite{Pa:72}, where the convergence of $N_n(\lambda)$ to the semicircle law was shown under the minimal conditions on
the distribution of $W=\big\{w_{jk}^{(n)}\big\}_{j,k=1}^{n}$ (the Lindeberg type conditions).

The second classical ensemble which we consider in the paper is  a sample covariance matrix of the form
\begin{equation}
M=n^{-1}XX^*,  \label{Msc}
\end{equation}
where $X$ is a $n\times m$ matrix whose entries $\big\{X_{jk}^{(n)}\big\}_{j=1,.,n,k=1,.,m}$ are independent random
variables, satisfying the conditions
\begin{equation}
\mathbf{E}\big\{X_{jk}^{(n)}\big\}=0,\quad \mathbf{E}\big\{(X_{jk}^{(n)})^{2}\big\}=1.
  \label{Xmom12}
\end{equation}%
Corresponding results on the convergence of normalized  linear eigenvalue statistics to  integrals with the
Marchenko-Pastur distribution were obtained in \cite{MP:67}.

 Central Limit Theorem (CLT) for fluctuations of linear eigenvalue statistics is  a natural second step in
 studies of the eigenvalue distribution of any ensemble of random matrices.
 That is why there are a lot of papers, devoted to
the proofs of CLT for different ensembles of random matrices (see
\cite{An-Ze:06,Ba-Si:04,Gu:02,Jo:98a,LP:09,Si-So:98a,So:00c,ST:10}). CLT for the  traces of resolvents for the
classical Wigner and sample covariance matrices    was proved by Girko in 1975 (see \cite{Gi:01} and references
therein), but the expression for the variance found by him was rather complicated. A simple expression for the
covariance of the resolvent traces for the Wigner matrix in the case $E\{(w_{ii}^{(n)})^2\}=2$ was found in
\cite{KKP:96}. CLT for polynomial test functions for some generalizations of the Wigner and sample covariance matrices
was proved in \cite{An-Ze:06} by using moment methods. CLT for real analytic test functions for the Wigner and sample
covariance matrices was established in \cite {Ba-Si:04} under additional assumptions that $E\{(w_{ii}^{(n)})^2\}=2$,
$E\{(w_{jk}^{(n)})^{4}\}=3E^2\{(w_{jk}^{(n)})^{2}\}=3$ (or $E\{(X_{jk}^{(n)})^{4}\}=3E^2\{(X_{jk}^{(n)})^{2}\}$ for the
model (\ref{Msc})). In the recent paper \cite{LP:09} CLT for the linear eigenvalue statistics of the Wigner and sample
covariance matrix ensemble was proved under assumptions that $E\{(w_{ii}^{(n)})^2\}=2$,  the third and the forth
moments of all entries are the same, but $E\{(w_{jk}^{(n)})^{4}\}$ is not necessary 3. Moreover,  the test functions,
studied in \cite{LP:09}, are not supposed to be real analytic. It was assumed that the Fourier transform
$\widehat\varphi$ of the test function $\varphi$ satisfies the inequality
\begin{equation}\label{condLP}
\int(1+|k|^5)|\widehat\varphi(k)|dk<\infty,
\end{equation}%
which means that $\varphi$ has more than 5 bounded derivatives.

In the present paper we prove CLT for the Wigner ensemble (\ref{MW}) under the following assumptions on the matrix
entries
\begin{equation}
 \mathbf{E}\big\{(w_{jk}^{(n)})^{4}\big\}=w_4, \;
 \; \sup_n\sup_{1\le j<k\le
n}\mathbf{E}\big\{|w_{jk}^{(n)}|^{4+\varepsilon_1}\big\}=w_{4+\varepsilon_1}<\infty, \; \varepsilon_1>0.
  \label{cond_w}
\end{equation}%
We consider the test functions from the  space $\mathcal{H}_s$, possessing the norm (cf (\ref{condLP}))
\begin{equation}\label{norm}
    ||\varphi||_s^2=\int(1+2|k|)^{2s}|\widehat\varphi(k)|^2dk,\quad s>3/2,\quad
    \widehat\varphi(k)=\frac{1}{2\pi}\int e^{ikx}\varphi(x)dx.
\end{equation}
\begin{theorem}\label{t:1}
Consider the Wigner model with entries satisfying condition (\ref{cond_w}). Let the real
valued test function $\varphi$ satisfy condition $||\varphi||_{3/2+\varepsilon}<\infty$ ($\varepsilon>0$).
Then $\mathcal{N}^\circ_n[\varphi]$ converges in distribution to the Gaussian random variable with
zero mean and the variance
\begin{align}\label{V}
V[\varphi ]=&\frac{1}{2 \pi ^{2}}\int_{-2}^{2}\int_{-2}^{2}\left( \frac{
\varphi(\lambda_1)-\varphi(\lambda_2)}{\lambda_1-\lambda_2 }\right) ^{2}\frac{4-\lambda _{1} \lambda
_{2}}{\sqrt{4-\lambda _{1}^{2}}\sqrt{4-\lambda _{2}^{2}}}d\lambda _{1}d\lambda _{2}
\\&+\frac{\kappa _{4}}{2\pi ^{2}}\left(
\int_{-2}^{2}\varphi (\mu )\frac{2-\mu ^{2}}{\sqrt{4-\mu ^{2}}}%
d\mu \right) ^{2}+\frac{w_2-2}{4\pi ^{2}}\left(
\int_{-2}^{2}\frac{\varphi (\mu )\mu }{\sqrt{4-\mu ^{2}}}%
d\mu \right) ^{2}, \notag
\end{align}
where $\kappa_4=w_4-3$.
\end{theorem}
Let us note that similarly to the result of \cite{LP:09} it is easy to check that Theorem \ref{t:1} remains valid if
the second condition of (\ref{cond_w}) is replaced by the Lindeberg type condition for the fourth moments of entries of
$W$
\begin{equation}
\lim_{n\rightarrow \infty }L_{n}^{(4)}(\tau )=0,\quad \forall \tau >0,
\label{Lin4}
\end{equation}
where
\begin{equation}\label{Ln}
L_{n}^{(4)}(\tau )=\frac{1}{n^{2}}\sum_{j,k=1}^{n}\mathbf{E}\{(w_{jk}^{(n)})^{4}
\mathbf{1}_{|w_{jk}^{(n)}|>\tau\sqrt n}\}.
\end{equation}
The proof will be the same as for Theorem \ref{t:1}, but  everywhere below   $n^{-\varepsilon_1/2}$ will be replaced by
$L_n(\tau)/\tau^\gamma$, with some positive $\gamma$.

The proof of Theorem \ref{t:1} is based on some combination of the resolvent approach with martingal bounds for the
variance of the resolvent traces, used before by many authors, in particularly, by Girko (see \cite{Gi:01} and
references therein). An important advantage of our approach is that it is shown by the marginal difference method that
(see Proposition \ref{p:b_var} below)
\begin{equation}\label{b_v}
\mathbf{Var}\{\mathrm{Tr\,}G(z)\}\le C/|\Im z|^4,\quad G(z)=(M-z)^{-1},\end{equation} while in the  previous papers the
martingal method was used only to obtain the bounds of the type $\mathbf{Var}\{\mathrm{Tr\,}G(z)^{-1}\}\le nC(z)$. The
bound (\ref{b_v}) will be combined with the inequality
\begin{proposition}\label{p:joh} For any $s>0$ and any $M$
\begin{equation}\label{pj.1}
\mathbf{Var}\{\mathcal{N}_n[\varphi]\}\le C_s||\varphi||_s^2\int_0^\infty dy
e^{-y}y^{2s-1}\int_{-\infty}^\infty\mathbf{Var}\{\mathrm{Tr\,}G(x+iy)\}dx.
\end{equation}
\end{proposition}
The proposition allows one to transform the bounds for the variances of the resolvent traces into the bounds for the
variances of linear eigenvalue statistics of $\varphi\in\mathcal{H}_s$, where the value of $s$ depends on the exponent
of $|\Im z|$ in the r.h.s. of (\ref{b_v}). It is important, that Proposition \ref{p:joh} has a rather general form and
therefore it is applicable to any ensemble of random matrices for which the bounds of the type (\ref{b_v}) (may be with
a different exponent of $|\Im z|$) are found. This makes Proposition \ref{p:joh} an important tool of the proof of CLT
for linear eigenvalue statistics for different random matrices. The idea of Proposition \ref{p:joh} was taken from the
paper \cite{Jo:98a}, where a similar argument was used to study the first order correction terms of
$n^{-1}\mathbf{E}\{\mathcal{N}_n[\varphi]\}$ for the matrix models. Having in mind Proposition \ref{p:joh}, one can
prove CLT for any dense in $\mathcal{H}_s$ set of the test functions, and then extend this result to the whole
$\mathcal{H}_s$ by the standard procedure (see Proposition \ref{p:CLTcont}). In the present paper for this aim we use
a set of convolutions of  integrable functions with the Poisson kernel (see (\ref{phi_e}) and (\ref{P_y})). This choice
simplifies considerably the argument in the proof of CLT and makes the proof  more short than that in the previous
papers \cite{An-Ze:06,Ba-Si:04,LP:09}.

The result for  sample covariance matrices is very similar. We assume that the moments of the entries of  $X$ from
(\ref{Msc}) satisfy the bounds
\begin{equation}
 \mathbf{E}\big\{(X_{jk}^{(n)})^{4}\big\}=X_4, \quad
\; \sup_n\sup_{1\le j<k\le n}\mathbf{E}\big\{|X_{jk}^{(n)}|^{4+\varepsilon_1}\big\}=X_{4+\varepsilon_1}<\infty, \;
\varepsilon_1>0.
  \label{cond_X}
\end{equation}%

\begin{theorem}\label{t:2}
Consider a random matrix (\ref{Msc}) -- (\ref{Xmom12}) with entries of $X$, satisfying the condition (\ref{cond_X}).
Let the real valued test function $\varphi$ satisfy condition $||\varphi||_{3/2+\varepsilon}<\infty$ ($\varepsilon>0$).
Then $\mathcal{N}^\circ_n[\varphi]$ in the limit $m,n\to\infty$, $m/n\to c\ge 1$ converges in distribution to the
Gaussian random variable with zero mean and the variance
\begin{align}\notag
V_{SC}[\varphi ]=&\frac{1}{2\pi ^{2}}\int_{a_{-}}^{a_{+}}%
\int_{a_{-}}^{a_{+}}\left( \frac{\Delta \varphi }{\Delta \lambda }\right)
^{2}  \frac{\Big(4c-(\lambda _{1}-a_{m})(\lambda _{2}-a_{m})\Big)d\lambda _{1}d\lambda _{2}}{\sqrt{%
4c-(\lambda _{1}-a_{m})^{2}}\sqrt{4c-(\lambda _{2}-a_{m})^{2}}}%
 \\&+\frac{\kappa _{4}}{4c\pi ^{2}}%
\left( \int_{a_{-}}^{a_{+}}\varphi (\mu )\frac{\mu -a_{m}}{\sqrt{%
4c-(\mu -a_{m})^{2}}}d\mu \right) ^{2},  \label{VSC}
\end{align}%
where  $\dfrac{\Delta \varphi }{\Delta \lambda }=\dfrac{\varphi(\lambda_1)-\varphi(\lambda_2)}{\lambda_1-\lambda_2}$,
$\kappa _{4}=X _{4}-3$ is the fourth cumulant of entries of $X$, $a_{\pm}=(1\pm\sqrt c)^2$, and
$a_m=\frac{1}{2}(a_++a_-)$.

\end{theorem}

\section{Proofs}
\textit{Proof of Proposition \ref{p:joh}.}  Consider the operator $\mathcal{D}_s$
\begin{equation}\label{D_s}
   \widehat{ \mathcal{D}_sf}(k)=(1+2|k|)^s\widehat f(k)
\end{equation}
It is easy to see that for fixed $n$ $\mathbf{Var}\{\mathcal{N}_n[\varphi]\}$ is a bounded quadratic form in the
Hilbert space $\mathcal{H}$ of the functions with the inner product $(u,v)_s=(\mathcal{D}_su,\mathcal{D}_sv)$, where
the symbol $(.,.)$ means the standard inner product of $L_2(\mathbb{R})$. Hence there exists a positive  self adjoint
operator $\mathcal{V}$ such that
\[\mathbf{Var}\{\mathcal{N}_n[\varphi]\}=(\mathcal{V}\varphi,\varphi)
=\hbox{Tr} (\Pi_\varphi\mathcal{V}\Pi_\varphi)
\]
where $\Pi_\varphi$ is the projection on the vector $\varphi$
\[ (\Pi_\varphi f)(x)=\varphi(x)(f,\varphi)||\varphi||_0^{-1}\]
where $||.||_0$ means the norm (\ref{norm}) with $s=0$. We can write
\[\hbox{Tr} (\Pi_\varphi\mathcal{V}\Pi_\varphi)=
\hbox{Tr} (\Pi_\varphi\mathcal{D}_s\mathcal{D}_s^{-1}\mathcal{V}\mathcal{D}_s^{-1}\mathcal{D}_s\Pi_\varphi)
\]
But it is easy to see that
\[(\mathcal{D}_s\Pi_\varphi f)(x)=(\mathcal{D}_s\varphi)(x)(f,\varphi)||\varphi||_0^{-1},\]
hence,
\[||\mathcal{D}_s\Pi_\varphi||=||\mathcal{D}_s\varphi||_0=||\varphi||_s.\]
Therefore we can write
\begin{equation}\label{pj.2}
\mathbf{Var}\{\mathcal{N}_n[\varphi]\}=\hbox{Tr} (\Pi_\varphi\mathcal{D}_s\mathcal{D}_s^{-1}
\mathcal{V}\mathcal{D}_s^{-1}\mathcal{D}_s\Pi_\varphi)\le||\mathcal{D}_s\Pi_\varphi||^2\hbox{Tr} (\mathcal{D}_s^{-1}
\mathcal{V}\mathcal{D}_s^{-1})
\end{equation}
But since for any $u,v\in L_2(\mathbb{R})$ we have
\begin{align*}
&\Gamma(2s)(\mathcal{D}_s^{-2}u,v)= \Gamma(2s)\int(1+2|k|)^{-2s}\widehat{ u}(k)\overline{\widehat{v}(k)}dk\\
&= \int_0^\infty dy e^{-y}y^{2s-1}\int e^{-2|k|y}\widehat{ u}(k)\overline{\widehat{v}(k)}dk
=\int_0^\infty dy
e^{-y}y^{2s-1}(P_y*u,P_y*v)\\
&= \int_0^\infty dy e^{-y}y^{2s-1}\int dx\int\int P_y(x-\lambda)P_y(x-\mu)u(\lambda)\overline{v(\mu)}d\lambda d\mu,
\end{align*}
where the symbol $*$ means the convolution of functions, and $P_y$ is the Poisson kernel
\begin{equation}\label{P_y}
    P_y(x)=\frac{y}{\pi(x^2+y^2)}.
\end{equation}
This implies
\begin{equation}\label{ker_D}
\Gamma(2s)\mathcal{D}_s^{-2}(\lambda,\mu)=\int_0^\infty dy e^{-y}y^{2s-1}\int dx P_y(x-\lambda)P_y(x-\mu),
\end{equation}
and so
\begin{align*}
\Gamma(2s)\hbox{Tr} (\mathcal{D}_s^{-1}\mathcal{V}\mathcal{D}_s^{-1})&= \int_0^\infty dy
e^{-y}y^{2s-1}\int dx\Big(\mathcal{V} P_y(x-.),P_y(x-.)\Big)\\
&= \int_0^\infty dy e^{-y}y^{2s-1}\int dx\mathbf{Var}\{\mathcal{N}_n[P_y(x-.)]\}
\\&=
\int_0^\infty dy e^{-y}y^{2s-1}\int dx\mathbf{Var}\{\Im\mathrm{Tr\,}G(x+iy)\}.
\end{align*}
This relation combined with (\ref{pj.2}) proves (\ref{pj.1}).$\square$

\medskip

In what wallows we  need to estimate $\mathbf{E}\big\{|w_{jk}^{(n)}|^{8}\big\}$ (see the proof of Proposition
\ref{p:b_var}). Hence, if $\varepsilon_1<4$, then it is convenient to consider the truncated matrix
\begin{equation}\label{trunc}
    \widetilde M^{(\tau)}=\{\widetilde M_{ij}^{(\tau)}\}_{i,j=1}^n,
    \quad \widetilde M_{ij}^{(\tau)}=M_{ij}\mathbf{1}_{|M_{ij}|\le
    \tau},\quad \widetilde M^{(\tau)\circ}-\mathbf{E}\{\widetilde M^{(\tau)}\}.
\end{equation}
\begin{lemma}\label{l:trunc} Let $\widetilde{\mathcal{N}}_n[\varphi]=\emph{Tr }\varphi(\widetilde M^{(\tau)\circ})$
be the linear eigenvalue statistic of the matrix $\widetilde M^{(\tau)\circ}$, corresponding to the test function
$\varphi$ with bounded first derivative. Then
\[|e^{ix\mathcal{N}_n^\circ[\varphi]}-e^{ix\widetilde{\mathcal{N}}_n[\varphi]}|\le o(1)+
C|x|\,||\varphi'||_\infty L_n(\tau)/\tau^3.\]
\end{lemma}
\textit{Proof}. Consider the matrix $M(t)=\widetilde M+t( M-\widetilde M)$. Let $\{\lambda_i(t)\}$ be eigenvalues of
$M(t)$ and $\{\psi_i(t)\}$ be corresponding eigenvectors. Then
\begin{align*}
&\mathbf{E}\Big\{\big|\mathcal{N}_n[\varphi]-\widetilde{\mathcal{N}}_n[\varphi]\big|\Big\}=\int_0^1
dt\mathbf{E}\Big\{\sum\big|\varphi'(\lambda_i(t))\lambda_i'(t)\big|\Big\}
\\&\le||\varphi'||_\infty\int_0^1 dt\mathbf{E}\Big\{\sum\big|(M'(t)\psi_i(t),\psi_i(t))\big|\Big\}
\le||\varphi'||_\infty\mathbf{E}\Big\{\hbox{Tr }|M-\widetilde
M|\Big\}\\&=||\varphi'||_\infty\sum_k\mathbf{E}\Big\{\Big|\sum_{ij}u_{kj}^*( M-\widetilde
M)_{ij}u_{jk}\Big|\Big\}\\
&\le||\varphi'||_\infty\mathbf{E}\Big\{\sum_{ij}\big|(M-\widetilde M)_{ij}\big|\Big\}\le\sup|\varphi'|L_n(\tau)/\tau^3,
\end{align*}
where $M'(t)=\frac{d}{dt}M(t)=( M-\widetilde M)$,  $U=\{u_{ik}\}$ is the unitary matrix such that $M-\widetilde
M=U^*\Lambda U$, where $\Lambda$ is a diagonal matrix and $|M-\widetilde M|=U^*|\Lambda |U$. Hence,
\begin{align*}
|e^{ix\mathcal{N}_n^\circ[\varphi]}-e^{ix\widetilde{\mathcal{N}}_n[\varphi]^\circ}|&\le 2\mathbf{Pr}\{\widetilde
M^{(\tau)}\not=M\}+|x|\Big(\mathbf{E}\{\mathcal{N}_n[\varphi]\}-
\mathbf{E}\{\widetilde{\mathcal{N}}_n[\varphi]\}\Big)\\&\le o(1)+C|x|\,||\varphi'||_\infty
L_n(\tau)/\tau^3.\end{align*} $\square$

It follows from Lemma \ref{l:trunc} that for our purposes it suffices to prove CLT for
$\widetilde{\mathcal{N}}^\circ_n[\varphi]$. Hence, starting from this point we will assume that $M$ is replaced by
$\widetilde M^{(\tau)o}$, but to simplify notations we will write $M$ instead of $\widetilde M^{(\tau)o}$ just assuming
below that the matrix entries of $W$ satisfy conditions
\begin{align}
& \mathbf{E}\{w_{jk}\}=0,\;\mathbf{E}\{w_{jk}^2\}=1+o(1),
(j\not=k),\;\mathbf{E}\{w_{jj}^2\}=w_2+o(1),\label{cond_w1}\\&
 \mathbf{E}\{w_{jk}^4\}=w_4+o(1),\notag\\
& \mathbf{E}\{|w_{jk}|^6\}\le w_{4+\varepsilon_1}n^{1-\varepsilon_1/2}, \quad\mathbf{E}\{|w_{jk}|^8\}\le
    w_{4+\varepsilon_1}n^{2-\varepsilon_1/2}.
\label{cond_w2}\end{align} Here and below we  omit also  the super index $(n)$ of  matrix entries $w_{jk}^{(n)}$ and
$X_{jk}^{(n)}$.

\begin{proposition}\label{p:b_var} If the conditions  (\ref{cond_w1})  are satisfied, then
for any $1>\delta>0$
\begin{align}\label{b_v.1}
    \mathbf{Var}\{\gamma_n\}\le Cn^{-1}\sum_{i=1}^{n} \mathbf{E}\{|G_{ii}(z)|^{1+\delta}\}/|\Im z|^{3+\delta},\quad
   \mathbf{Var}\{\gamma_n\}\le  C/|\Im z|^4.
\end{align}
If the  conditions of (\ref{cond_w2}) are also satisfied, then
\begin{align}\label{b_v.1a}
\mathbf{E}\{|\gamma_n^\circ|^4\}\le  Cn^{-1-\varepsilon_1/2}/|\Im z|^{12}.
\end{align}
\end{proposition}
\textit{Proof.} Denote $\mathbf{E}_{\le k}$ the averaging with respect to $\{w_{ij}\}_{1\le i\le j\le k}$. Then,
according to the standard martingal method (see \cite{Dh-Co:68}), we have
\begin{align}\label{mart}\mathbf{Var}\{\gamma_n\}=\sum_{k=1}^n\mathbf{E}\{|\mathbf{E}_{\le k-1}\{\gamma_n\}-
\mathbf{E}_{\le k}\{\gamma_n\}|^2\}.\end{align}
Denote  $\mathbf{E}_{ k}$ the averaging with respect to
$\{w_{ki}\}_{1\le i\le n}$. Then, using the Schwarz inequality, we obtain that
\[|\mathbf{E}_{\le k-1}\{\gamma_n\}-\mathbf{E}_{\le k}\{\gamma_n\}|^2=
|\mathbf{E}_{\le k-1}\{\gamma_n-\mathbf{E}_{ k}\{\gamma_n\}|\le
\mathbf{E}_{\le k-1}\{|\gamma_n-E_{ k}\{\gamma_n\}|^2\}.
\]
Hence
\begin{equation}\label{b_v.2}
    \mathbf{Var}\{\gamma_n\}\le\sum_{k=1}^n\mathbf{E}\{|\gamma_n-\mathbf{E}_{ k}\{\gamma_n\}|^2\}.
\end{equation}
Let us estimate the first summand (with $k=1$) of the above sum. The other ones
can be estimated similarly. Denote $M^{(1)}$ the $(n-1)\times(n-1)$ matrix which is the main bottom $(n-1)\times(n-1)$
minor of $M$
\begin{equation}\label{G^1}
    G^{(1)}=(M^{(1)}-z)^{-1},\quad m^{(1)}=n^{-1/2}(w_{12},\dots,w_{1n})\in\mathbb{R}^{n-1}.
\end{equation}
We will use the  identities
\begin{align}&\hbox{Tr }G-\hbox{Tr }G^{(1)}= -\frac{1+(G^{(1)}G^{(1)}m^{(1)},m^{(1)})}{z+n^{-1/2}w_{11}+(G^{(1)}m^{(1)},m^{(1)})}
=:-\frac{1+B(z)}{A(z)}. \label{repr}\\
\notag &G_{11}=-A^{-1},\;\quad G_{ii}-G^{(1)}_{ii}=-(G^{(1)}m^{(1)})_i^2/A,
\end{align} where $(.,.)$ means the standard inner product in $\mathbb{C}^{n-1}$.

The first identity of (\ref{repr}) yields that  it suffices to estimate
$\mathbf{E}\{|BA^{-1}-\mathbf{E}_{1}\{BA^{-1}\}|^2\}$ and $\mathbf{E}\{|A^{-1}-\mathbf{E}_{1}\{A^{-1}\}|^2\}$. We will
estimate the first expression. The second one can be estimated similarly. Denote
$\xi^\circ_1=\xi-\mathbf{E}_{1}\{\xi\}$ for any random variable $\xi$ and note that for any $a$ independent of
$\{w_{1i}$ we have
\[\mathbf{E}_1\{|\xi^\circ_1|^2\}\le \mathbf{E}_1\{|\xi-a|^2\}. \]
Hence it suffices to estimate
\[\bigg|\frac{B}{A}-\frac{\mathbf{E}_{1}\{B\}}{\mathbf{E}_{1}\{A\}}\bigg|=
\bigg|\frac{B^\circ_1}{\mathbf{E}_{1}\{A\}}-
\frac{A^\circ_1}{\mathbf{E}_{1}\{A\}}\,\frac{B}{A}\bigg|\le
\bigg|\frac{B^\circ_1}{\mathbf{E}_{1}\{A\}}\bigg|+
\bigg|\frac{A^\circ_1}{\Im z\mathbf{E}_{1}\{A\}}\bigg|
\]
Let us use also the identities that follow from the spectral theorem
\begin{equation}\label{sp_rel}
\Im(G^{(1)}m^{(1)},m^{(1)})=\Im z(|G^{(1)}|^2m^{(1)},m^{(1)}),\quad \Im\hbox{Tr }G^{(1)}=\Im z\hbox{Tr }|G^{(1)}|^2.
\end{equation}
where $|G^{(1)}|=(G^{(1)}G^{(1)*})^{1/2}$. The first relation  yields, in particular, that $|{B}/{A}|\le|\Im z|^{-1}$.
Moreover, using the second identity  of (\ref{sp_rel}), we have
\begin{equation}\label{b_v.2a}
\frac{n^{-1}\hbox{Tr }|G^{(1)}|^2}{|z+n^{-1}\hbox{Tr }G^{(1)}|^2}= \frac{(n^{-1}\hbox{Tr
}|G^{(1)}|^{2})^{\delta}(n^{-1}\hbox{Tr }|G^{(1)}|^{2})^{1-\delta}} {|z+n^{-1}\hbox{Tr
}G^{(1)}|^{1+\delta}|z+n^{-1}\hbox{Tr }G^{(1)}|^{1-\delta}} \le C\frac{|\Im
z|^{-1-\delta}}{|\mathbf{E}_{1}\{A\}|^{1+\delta}}.\end{equation} Since
\begin{align}\label{A^0}
&A^\circ_1=n^{-1/2}w_{11}+n^{-1}\sum_{i\not=j}G^{(1)}_{ij}w_{1i}w_{1j}+n^{-1}\sum_{i}G^{(1)}_{ii}(w_{1i}^2)^\circ,\\
&\mathbf{E}_1\{|A^\circ_1|^2\}\le Cn^{-2}\hbox{Tr }|G^{(1)}|^2+Cn^{-1},\notag\end{align} we get by (\ref{b_v.2a}) and
the second identity of (\ref{sp_rel}):
\begin{equation}\label{b_v.2b}
\mathbf{E}_1\Big\{\Big|\frac{A^\circ_1}{\mathbf{E}_{1}\{A\}}\Big|^2\Big\}\le C\big(|\Im
z||\mathbf{E}_{1}\{A\}|\big)^{-1-\delta}.\end{equation}
 Similarly
\[\mathbf{E}_1\bigg\{\bigg|\frac{B^\circ_1}{\mathbf{E}_{1}\{A\}}\bigg|^2\bigg\}\le
\frac{Cn^{-2}\hbox{Tr }|G^{(1)}|^4}{|z+n^{-1}\hbox{Tr }G^{(1)}|^2}\le \frac{C|\Im z|^{-2}n^{-2}\hbox{Tr
}|G^{(1)}|^2}{|z+n^{-1}\hbox{Tr }G^{(1)}|^2}\le C\frac{n^{-1}|\Im z|^{-3-\delta}}{|\mathbf{E}_{1}\{A\}|^{1+\delta}}.\]
Then, using  the Jensen inequality $|\mathbf{E}_1\{A\}|^{-1}\le\mathbf{E}_1\{|A|^{-1}\}$, and the second identity of
(\ref{repr}), we conclude that
\[\mathbf{E}\{|(\gamma_n(z))^\circ_1|^2\}\le \frac{C}{n|\Im z|^{3+\delta}}\mathbf{E}\{|G_{11}(z)|^{1+\delta}\}.\]
Then (\ref{b_v.2}) implies (\ref{b_v.1}).

To prove  (\ref{b_v.1a}) we use the inequality similar to (\ref{b_v.2}) (see \cite{Dh-Co:68})
\begin{equation}\label{b_v.3}
    \mathbf{E}\{|\gamma_n^\circ|^4\}\le Cn\sum_{k=1}^n\mathbf{E}\{|\gamma_n-\mathbf{E}_{ k}\{\gamma_n\}|^4\}.
\end{equation}
Thus, in view of (\ref{repr}), it is enough to check that
\begin{equation}\label{b_v.4}
    \mathbf{E}_1\{|A^\circ_1|^4\}\le Cn^{-1-\varepsilon_1/2}|\Im z|^{-4},\quad
\mathbf{E}_1\{|B^\circ|^4_1\}\le Cn^{-1-\varepsilon_1/2}|\Im z|^{-8}.
\end{equation}
The first relation here evidently follow from (\ref{A^0}), if we take the forth degree of the r.h.s., average with
respect to $\{w_{1i}\}$, and take into account (\ref{cond_w2}).  The second relation can be obtained similarly.
$\square$

Proposition \ref{p:b_var} gives the bound for the variance of the linear eigenvalue statistics for the functions
$\varphi(\lambda)=(\lambda-z)^{-1}$. We are going to extend the bound for a  wider class of test functions.

\begin{lemma}\label{l:2} If $||\varphi||_{3/2+\epsilon}\le\infty$, with any $\epsilon>0$, then
\begin{equation}\label{l2.1}
   \mathbf{ Var}\{\mathcal{N}_n[\varphi]\}\le C_\varepsilon||\varphi||_{3/2+\varepsilon}^2.
\end{equation}
\end{lemma}
\textit{Proof.}
In view of Proposition
\ref{p:joh} we need to estimate
\[I(y)=\int_{-\infty}^{\infty}\mathbf{Var}\{\gamma_n(x+iy)\}dx
\]
Take in (\ref{b_v.1})  $\delta=\varepsilon/2$. Then  we need to estimate
\[\int_{-\infty}^{\infty}\mathbf{E}\{|G_{jj}(x+iy)|^{1+\varepsilon/2}\}dx,\,j=1,\dots,n.
\]
We do this for $j=1$. For other $j$ the estimates are the same. The spectral representation
\[{G}_{11}=\int\frac{N_{11}(d\lambda)}{\lambda-x-iy}\]
 and the Jensen inequality yield
\[\int_{-\infty}^{\infty} |{G}|_{11}^{1+\varepsilon/2}(x+iy)dx\le\int_{-\infty}^{\infty}dx
\int_{-\infty}^{\infty}\frac{N_{11}(d\lambda)}{(|x-\lambda|^2+y^2)^{(1+\varepsilon/2)/2}} \le C|y|^{-\varepsilon/2}.\]
Taking $s=3/2+\varepsilon$ in (\ref{pj.1}), we get
\[ \mathbf{ Var}\{\mathcal{N}_n[\varphi]\}\le ||\varphi||_{3/2+\varepsilon}^2C
\int_0^\infty
e^{-y}y^{2+2\varepsilon}y^{-3-\varepsilon}dy\le C||\varphi||_{3/2+\varepsilon}^2.\]
 $\square$

To simplify formulas we will assume below that $\{w_{jk}\}_{1\le j<k\le n}$ are i.i.d. and $\{w_{jj}\}_{1\le j\le n}$
are i.i.d. Note that this assumption does not change the proof seriously, it just allows us to write the bounds only
for $G_{11}$ instead of all $G_{ii}$.

 The next lemma collects relations which we need  to prove CLT.
\begin{lemma}\label{l:*} Using notations of (\ref{repr}) we have uniformly in $z_1,z_2:\Im z_{1,2}>a$
with any $a>0$:
\begin{align}\label{*.1}
&\mathbf{E}\{(A^\circ)^3\},\;\mathbf{E}\{|A^\circ|^4\},\;\mathbf{E}\{(B^\circ)^3\},\;\mathbf{E}\{|B^\circ|^4\}
=O(n^{-1-\varepsilon_1/2}),\\
& n\mathbf{E}_1\{A^\circ(z_1)A^\circ(z_2)\}=\frac{2}{n}\mathrm{Tr }G^{(1)}(z_1)G^{(1)}(z_2)+w_2\label{*.1a}\\
& \hskip3cm +\frac{\kappa_4}{n} \sum_{i}G^{(1)}_{ii}(z_1)G^{(1)}_{ii}(z_2)+\overset\circ{\gamma}_n^{(1)}(z_1)
\overset\circ{\gamma}_n^{(1)}(z_2)/n, \notag\\
& n\mathbf{E}_1\{A^\circ(z_1)B^\circ(z_2)\}=n\frac{d}{dz_2}\mathbf{E}_1\{A^\circ(z_1)A^\circ(z_2)\},\label{*.1b}
\end{align}
\begin{align}
\mathbf{Var}\{n\mathbf{E}_1\{A^{\circ}(z_1)A^{\circ}(z_2)\}\}&= O(n^{-1}),\label{*.1c} \\
\mathbf{Var}\{n\mathbf{E}_1\{A^{\circ}(z_1)B^{\circ}(z_2)\}\}&=O(n^{-1}),\notag\\
\mathbf{E}\{|\overset\circ\gamma_n^{(1)}(z)-\overset\circ\gamma_n(z)|^4\}&=  O(n^{-1-\varepsilon_1/2}).\label{*.1d}
\end{align}
Moreover,
\begin{align}\label{*.2}
& \mathbf{Var}\{G^{(1)}_{ii}(z_1)\}= O(n^{-1}),\quad
|\mathbf{E}\{G^{(1)}_{ii}(z_1)\}-\mathbf{E}\{G_{ii}(z_1)\}|= O(n^{-1})\\
& |\mathbf{E}\{\gamma_n^{(1)}(z)\}/n-f(z)|= O(n^{-1}),\quad|\mathbf{E}^{-1}\{A(z)\}+f(z)|= O(n^{-1}).\label{*.2a}
\end{align}
\end{lemma}
\textit{Proof.} Note that since $\Im z\Im (G^{(1)}m,m)\ge 0$, we can use the bound
\begin{equation}\label{b_A}|\Im A|\ge|\Im z|\Rightarrow |A^{-1}|\le|\Im z|^{-1}\le a^{-1}.\end{equation}
Relations (\ref{*.1}) follow from the representations
\begin{align}\label{*.3}
A^\circ&= A^\circ_1+n^{-1} \overset\circ\gamma_n^{(1)}(z),\quad B^\circ=B^\circ_1 +n^{-1}
\frac{d}{dz}\overset\circ\gamma_n^{(1)}(z),\end{align} combined with (\ref{b_v.4}),  and   (\ref{b_v.1a}), applied to
$\gamma_n^{(1)}$. Relations (\ref{*.1a}) and (\ref{*.1b}) follow  from (\ref{A^0}) and (\ref{*.3}), if we take the
products of the r.h.s. of (\ref{A^0}) with different $z$ and average with respect to $\{w_{1i}\}$. Relation
(\ref{*.1d}) follows from (\ref{repr}), (\ref{*.1}), and (\ref{b_A}). The first relation of (\ref{*.2}) is the analog
of the relation
\begin{equation}\label{*.3a}
\mathbf{Var}\{G_{ii}(z_1)\}=\mathbf{Var}\{G_{11}(z_1)\}= O(n^{-1})\end{equation} if in the latter we replace the matrix
$M$ by $M^{(1)}$. But since $G_{11}(z_1)=-A^{-1}(z_1)$, (\ref{*.3a}) follows from (\ref{*.1}) and  (\ref{b_A}).  The
second relation of (\ref{*.2}) follows from  (\ref{repr}).

The first relations of (\ref{*.2a}) follows from the above bound for $n^{-1}\mathbf{E}\{\gamma_n-\gamma_n^{(1)}\}$ and
the well known estimate (see e.g. \cite{KKP:96})
\[n^{-1}\mathbf{E}\{\gamma_n\}-f(z)=O(n^{-1}).\]
The second one of (\ref{*.2a}) is the corollary of the above estimate and of the relation
\[\mathbf{E}^{-1}\{A(z)\}=(z+\mathbf{E}\{\gamma_n^{(1)}\}/n)^{-1}=(z+f(z))^{-1}+O(n^{-1})=-f(z)
+O(n^{-1}).\]

 Finally we obtain the first bound of (\ref{*.1c}) from (\ref{*.1a}), (\ref{*.2}), (\ref{*.1d}),
 and the identity
\begin{equation}\label{res_id}
\mathrm{Tr\,}G^{(1)}(z_1)G^{(1)}(z_2)=\mathrm{Tr\,}\frac{G^{(1)}(z_1)-G^{(1)}(z_2)}{z_1-z_2}.\end{equation} The second
bound of (\ref{*.1c}) follows from the first one, (\ref{*.1b}), and the Cauchy theorem. $\square$

\bigskip

\textit{Proof of Theorem \ref{t:1}}. We prove first Theorem \ref{t:1} for the functions
$\varphi_\eta$ of the form
\begin{equation}\label{phi_e}
\varphi_\eta=P_\eta*\varphi_0,\quad \int|\varphi_0(\lambda)|d\lambda\le C<\infty,
\end{equation}
where $P_\eta$ is the Poisson kernel defined in (\ref{P_y}). One can see easily that
\begin{equation}\label{repr_N}
\mathcal{N}_n^\circ[\varphi]=\frac{1}{\pi}\int \varphi_0(\mu)\Im\gamma_n^\circ(z_\mu)d\mu,\quad
z_\mu=\mu+i\eta.
\end{equation}
Set
\begin{equation}\label{Z_n}
Z_n(x)=\mathbf{E}\{e^{ix\mathcal{N}_n^\circ[\varphi]}\},\quad e(x)=e^{ix\mathcal{N}_n^\circ[\varphi]}, \quad
Y_n(z,x)=\mathbf{E}\{\hbox{Tr}G(z)e^\circ(x)\}.
\end{equation}
Then
\begin{align}\label{CLT.1}
\frac{d}{dx}Z_n(x)&= \frac{x}{2\pi}\int \varphi_0(\mu)( Y(z_\mu,x)- Y(\overline z_\mu,x))d\mu.
\end{align}
On the other hand, using the symmetry of the problem and
the notations of (\ref{repr}), we have
\begin{align}\label{Y=}
Y_n(z,x)&= \mathbf{E}\{\hbox{Tr }G(z)e^\circ(x)\}=n\mathbf{E}\{G_{11}(z)e^\circ(x)\}\\
&=  -n\mathbf{E}\{(A^{-1})^\circ e_1(x)\}-n\mathbf{E}\{(A^{-1})^\circ (e(x)-e_1(x))\}=T_1+T_2, \notag\end{align} where
\[e_1(x)=e^{ix(\mathcal{N}_{n-1}^{(1)}[\varphi])^\circ}, \quad
(\mathcal{N}_{n-1}^{(1)}[\varphi])^\circ=(\hbox{Tr}\varphi(M^{(1)}))^\circ =\int
d\mu\,\varphi_0(\mu)\Im\overset\circ\gamma_n^{(1)}(z_\mu).\] Let us use the representation
\begin{equation}\label{A^-1}
    A^{-1}=\frac{1}{\mathbf{E}\{A\}}-\frac{A^\circ}{\mathbf{E}^{2}\{A\}}+\frac{(A^\circ)^2}{\mathbf{E}^{3}\{A\}}-
\frac{(A^\circ)^3}{\mathbf{E}^{4}\{A\}}+\frac{(A^\circ)^4}{A\mathbf{E}^{4}\{A\}}.
\end{equation}
Since $e_1(x)$ does not depend on $\{w_{1i}\}$, using that $\mathbf{E}\{...\}=\mathbf{E}\{\mathbf{E}_1\{...\}\}$, we
obtain in view of (\ref{A^-1}) and  (\ref{*.1})
\begin{align*}T_1&= \frac{\mathbf{E}\{
n\mathbf{E}_1\{A^{\circ}(z))\} e_1^\circ(x)\}}{\mathbf{E}^{2}\{A\}}-
\frac{\mathbf{E}\{n\mathbf{E}_1\{(A^{\circ}(z))^2\} e_1^\circ(x)\}}{\mathbf{E}^{3}\{A\}}+O(n^{-\varepsilon_1/2}).
\end{align*}
Relations  (\ref{*.1c})  implies
\[|\mathbf{E}\{n\mathbf{E}_1\{(A^{\circ}(z))^2\} e_1^\circ(x)\}|\le
\mathbf{Var}^{1/2}\{n\mathbf{E}_1\{(A^{\circ}(z))^2\}\}\mathbf{Var}^{1/2}\{e_1^\circ(x)\}=O(n^{-1/2}),\]
thus
\begin{align*}
T_1&= \mathbf{E}^{-2}\{A\}\mathbf{E}\{ \gamma^{(1)}_n e_1^\circ(x)\}+O(n^{-\varepsilon_1/2}).
\end{align*}
But the Schwarz inequality and (\ref{*.1d}) yield
\begin{align}\label{e-e_1}
|\mathbf{E}\{(\gamma^{(1)}_n)^\circ e_1(x)\}-\mathbf{E}\{\gamma_n^\circ e(x)\}|\le
\mathbf{E}\{|(\gamma^{(1)}_n)^\circ-\gamma_n^\circ|(1+|x||\gamma_n^\circ|)\}\\
\le C(1+|x|)\mathbf{E}^{1/2}\{|(\gamma^{(1)}_n)^\circ-\gamma_n^\circ|^2\}=O(n^{-1/2}).\notag
\end{align}
Thus,  we have
\begin{align}\label{T_1}
T_1=\mathbf{E}^{-2}\{A(z)\}Y_n(z,x)+O(n^{-1/2})=f^2(z)Y_n(z,x)+O(n^{-\varepsilon_1/2}).\end{align} To find $T_2$, we
write
\[e(x)-e_1(x)=ix\int\varphi_0(\mu)\Big(\Im(\gamma_n^\circ-\overset\circ\gamma^{(1)}_n)e_1(x)
+O((\gamma_n^\circ-(\gamma^{(1)}_n)^\circ)^2)\Big)d\mu.\] Using the Schwartz inequality, (\ref{repr}), (\ref{*.1d}),
and (\ref{*.1a}), we conclude that the term $O((\gamma_n^\circ-(\gamma^{(1)}_n)^\circ)^2)$ gives the contribution
$O(n^{-\varepsilon_1/4})$. Then, since $e_1(x)$ does not depend on $\{w_{1i}\}$, we  average first with respect to
$\{w_{1i}\}$ and obtain in view of (\ref{*.1d})
\begin{align*}
&T_2= -\frac{ix n}{\pi}\int d\mu\varphi_0(\mu)\mathbf{E}\Big\{e_1(x)(A^{-1})^\circ(z)
\Im\big(\gamma_n-\gamma^{(1)}_n\big)^\circ(z_\mu)\Big\}+O(n^{-\varepsilon_1/4})\\
&= \frac{ix n}{\pi}\int d\mu\varphi_0(\mu)\mathbf{E}\bigg\{e_1(x)\mathbf{E}_1\Big\{(A^{-1})^\circ(z)
\Im\Big(\frac{1+B(z_\mu)}{A(z_\mu)}\Big)^\circ\Big\}\bigg\}+O(n^{-\varepsilon_1/4})\\
&= \frac{ix n}{\pi}\int d\mu\varphi_0(\mu)\mathbf{E}\Big\{(A^{-1})^\circ(z)
\Im\Big(\frac{1+B(z_\mu)}{A(z_\mu)}\Big)^\circ\Big\}\mathbf{E}\{e_1(x)\}+O(n^{-\varepsilon_1/4}).
\end{align*}
Using (\ref{A^-1}) and (\ref{*.1}), we conclude that only linear terms with respect to $B^\circ$
and $A^\circ$  give non vanishing contribution, hence in view of (\ref{*.1b}) and (\ref{*.1c})
we obtain
\begin{align*}&D_n(z,z_\mu)
:=n\mathbf{E}\Big\{(A^{-1})^\circ(z)\Big((1+B(z_\mu))A^{-1}(z_\mu)\Big)^\circ\Big\}\\
&=\frac{\big(1+\mathbf{E}\{B(z_\mu)\}\big)n\mathbf{E}_1\{A^\circ(z)A^\circ(z_\mu)\}}
{ \mathbf{E}^{2}\{A(z)\}\mathbf{E}^{2}\{A(z_\mu)\}}-\frac{n\mathbf{E}_1\{A^\circ(z)B^\circ(z_\mu)\}}{\mathbf{E}^{2}\{A(z)\}\mathbf{E}\{A(z_\mu)\}}+O(n^{-\varepsilon_1/2})\\
&= f^2(z)f^2(z_\mu)(1+f'(z_\mu))\Big(\mathbf{E}\Big\{\frac{2}{n}\mathrm{Tr
}G^{(1)}(z_1)G^{(1)}(z_2)\Big\}\\&+\frac{\kappa_4}{n}\mathbf{E}\Big\{
\sum_{i}G^{(1)}_{ii}(z_1)G^{(1)}_{ii}(z_2)\Big\}+2\Big)\\&
-f^2(z)f(z_\mu)\frac{d}{dz_\mu}\Big(\mathbf{E}\Big\{\frac{2}{n}\mathrm{Tr
}G^{(1)}(z)G^{(1)}(z_\mu)\Big\}\\&+\frac{\kappa_4}{n}\mathbf{E}\Big\{
\sum_{i}G^{(1)}_{ii}(z)G^{(1)}_{ii}(z_\mu)\Big\}\Big)+O(n^{-\varepsilon_1/2}),
\end{align*}
where we used also (\ref{*.2a}) to replace $\mathbf{E}^{-1}\{A(z)\}$ by $f(z)$ and $\mathbf{E}\{B(z_\mu)\}$ by
$f'(z_\mu)$. The identity (\ref{res_id})  yields
\begin{align}\label{D_n}
D_n(z,z_\mu):=&2f^2(z)f^2(z_\mu)(1+f'(z_\mu))\Big(\frac{f(z)-f(z_\mu)}{z-z_\mu}+w_2/2\Big)\\
& +2f^2(z)f(z_\mu)\frac{d}{dz_\mu}\Big(\frac{f(z)-f(z_\mu)}{z-z_\mu}\Big)\notag\\
& + \kappa_4\Big(f^3(z)f^3(z_\mu)(1+f'(z_\mu))+f^3(z)f(z_\mu)f'(z_\mu)\Big)+O(n^{-\varepsilon_1/2}).\notag
\end{align}
In addition, similarly to (\ref{e-e_1}), we have
\[\mathbf{E}\{e_1(x)\}=Z_n(x)+O(n^{-1/2}).\]
Hence, relations (\ref{Y=})--(\ref{D_n}) imply
\begin{align}\notag
& Y_n(z,x)=f^2(z)Y_n(z,x)+ixZ_n(x)\int d\mu\varphi (\mu)\frac{D_n(z,z_\mu)-D_n(z,\overline{z_\mu})}{2i\pi}+o(1),\\
& Y_n(z,x)=ixZ_n(x)\int d\mu\varphi_0 (\mu)\frac{C_n(z,z_\mu)-C_n(z,\overline{z_\mu})}{2i\pi}+o(1),\label{Y} \\
&  C_n(z,z_\mu):=\frac{D_n(z,z_\mu)}{1-f^2(z)}. \notag\end{align}
Using the relations
\[
f(z)(f'(z)+1)=\frac{f(z)}{1-f^2(z)}=-\frac{1}{\sqrt{z^2-4}},\quad
f'=-\frac{f(z)}{\sqrt{z^2-4}},\]
we can transform $C_n(z,z_\mu)$ to the form
\begin{align}\notag
C_n(z,z_\mu)&= \frac{1}{(z-z_\mu)^2}
\bigg(\frac{zz_\mu-4}{(z^2-4)^{1/2}(z_\mu^2-4)^{1/2}}-1\bigg)+\frac{(w_2-2)f(z)f(z_\mu)}
{(z^2-4)^{1/2}(z_\mu^2-4)^{1/2}}\\&  +2\kappa_4\frac{f^2(z)f^2(z_\mu)}{(z^2-4)^{1/2}(z_\mu^2-4)^{1/2}}+o(1)=:
C(z,z_\mu)+o(1). \label{C}\end{align} Now, taking into account (\ref{CLT.1}), (\ref{Y}), and (\ref{C}), we obtain the
equation
\begin{align}\label{equ}
\frac{d}{dx}Z_n(x)&= -xV[\varphi_0,\eta]Z_n(x)+o(1)\\
V[\varphi_0,\eta]&= \frac{1}{4\pi^2}\int\int\varphi_0(\mu_1)\varphi_0(\mu_2)\Big(
C(z_{\mu_1},\overline{z_{\mu_2}})+C(\overline{z_{\mu_1}},z_{\mu_2})\notag\\& -
C(z_{\mu_1},z_{\mu_2})-C(\overline{z_{\mu_1}},\overline{z_{\mu_2}})\Big)d\mu_1 d\mu_2. \notag\end{align} Now if we
consider
\[\widetilde Z_n(x)=e^{x^2V[\varphi_0,\eta]/2}Z_n(x),\]
then (\ref{equ}) yields  for any $|x|\le C$
\[\frac{d}{dx}\widetilde Z_n(x)=o(1),\]
and since $\widetilde Z_n(0)=Z_n(0)=1$, we obtain uniformly in $|x|\le C$
\begin{align}\notag
& \widetilde Z_n(x)=1+o(1)\\
 \Rightarrow
 & Z_n(x)=e^{-x^2V[\varphi_\eta]/2}+o(1).
\label{conv_Z}\end{align} Thus, we have proved CLT for the functions of the form (\ref{phi_e}). To extend CLT to a
wider class of functions we use

\begin{proposition} \label{p:CLTcont}
Let $\{\xi_{l}^{(n)}\}_{l=1}^{n}$ be a triangular array of random variables,
$\displaystyle\mathcal{N}_{n}[\varphi ]=\sum_{l=1}^{n}\varphi
(\xi _{l}^{(n)})$ be its linear statistics,
corresponding to a test function $\varphi :\mathbb{R}\rightarrow
\mathbb{R}$, and
\[V_{n}[\varphi]=\mathbf{Var}\{\mathcal{N}_{n}[\varphi ]\}\]
be the variance of $\mathcal{N}_{n}[\varphi ]$. Assume that

(a) there exists a vector space $\mathcal{L}$ endowed with a norm $||...||$
and such that $V_{n}$ is defined on $\mathcal{L}$ and admits the bound
\begin{equation}
V_{n}[\varphi ]\leq C||\varphi ||^2,\;\forall \varphi \in \mathcal{L},
\label{ocVn1}
\end{equation}%
where $C$ does not depend on $n$;

(b) there exists a dense linear manifold $\mathcal{L}_{1}\subset \mathcal{L}$ such that  CLT is valid for
$\mathcal{N}_{n}[\varphi ],\;\varphi \in \mathcal{L}_{1}$, i.e., if $ Z_{n}[x\varphi ]=\mathbf{E}\Big\{
e^{ix\overset{\circ}{\mathcal{N}}_{n}[\varphi ]}\Big\} $ is the characteristic function of $n^{-1/2}\overset{\circ
}{\mathcal{N}}_{n}[\varphi ]$, then there exists a continuous quadratic functional $V:\mathcal{L}_{1}\rightarrow
\mathbb{R}_{+}$ such that we have uniformly in $x$, varying on any compact interval
\begin{equation}\label{limZ}
\lim_{n\rightarrow \infty }Z_{n}[x\varphi ]=e^{-x^{2}V[\varphi
]/2},\;\forall \varphi \in \mathcal{L}_{1};
\end{equation}
Then  $V$ admits a continuous extension to $\mathcal{L}$ and CLT is valid for all $\mathcal{N}_{n}[\varphi ]$,
 $\varphi \in \mathcal{L}$.
\end{proposition}

\textit{Proof.} Let $\{\varphi _{k}\}$ be a sequence of elements of
$\mathcal{L}_{1}$ converging to $\varphi \in \mathcal{L}$. We have
then in view of the inequality $ |e^{ia}-e^{ib}|\leq |a-b|$, the
linearity of $\overset{\circ}{\mathcal{N}} _{n}[\varphi ]$ in
$\varphi $, the Schwarz inequality, and (\ref{ocVn1}):
\begin{align*} \Big|Z_{n}(x\varphi
)-Z_{n}(x\varphi)|_{\varphi=\varphi_k} \Big| &\leq |x|\mathbf{E}\left\{ \left\vert
\overset{\circ}{\mathcal{N}}_{n}[\varphi ]-\overset{\circ}{ \mathcal{N}}_{n}[\varphi _{k}]\right\vert \right\} \\ &\leq
|x|\mathbf{Var}^{1/2}\{\mathcal{N}_{n}[\varphi -\varphi _{k}]\}\leq C|x|\;||\varphi -\varphi _{k}||.
\end{align*}
Now,
passing first to the limit $n\rightarrow \infty $ and then
$k\rightarrow \infty $, we obtain the assertion. $\square$

\medskip

The proposition and Lemma \ref{l:2} allow us to complete the proof of Theorem \ref{t:1}.$\square$

\medskip

\textit{Proof of Theorem \ref{t:2}} The proof of Theorem \ref{t:2} can be performed by the same way as that for Theorem
\ref{t:1}. We start from the proposition which is the analog of Proposition \ref{p:b_var}.
\begin{proposition}\label{p:b_var_MP} Let
$\gamma_n=\mathrm{Tr\,}G(z)$, where $G(z)=(M-z)^{-1}$ and $M$ is a sample covariance matrix (\ref{Msc}) with entries
satisfying (\ref{Xmom12}) and (\ref{cond_X}). Then inequalities (\ref{b_v.1}) hold.
\end{proposition}
Taking into account Proposition \ref{p:b_var_MP}, on the basis of Proposition \ref{p:joh} and Lemma \ref{l:2} we obtain
immediately the bound (\ref{l2.1}) for the variance of  linear eigenvalue statistics of  sample covariance matrices.
Then one can use the same method as in the proof of Theorem \ref{t:1} to prove CLT for $\varphi_\eta$ of (\ref{phi_e})
or just use the result of \cite{LP:09} for the functions, satisfying conditions (\ref{condLP}). Then Proposition
\ref{p:CLTcont} implies immediately the assertion of Theorem \ref{t:2}.

Thus, to complete the proof of Theorem \ref{t:2} we are left  to prove Proposition \ref{p:b_var_MP}.

\medskip

\textit{Proof of Proposition \ref{p:b_var_MP}}. Similarly to the proof of Proposition \ref{p:b_var} we use the identity
(\ref{mart}) where this time $\mathbf{E}_{\le k}$ means the averaging with respect to $\{X_{jl}\}_{l=1,.,m,j\le k}$.
Then we obtain (\ref{b_v.2}) with $\mathbf{E}_{ k}$ meaning the averaging with respect to $\{X_{kl}\}_{l=1,.,m}$.

Denote $M^{(1)}=X^{(1)}X^{(1)*}$, where the $(n-1)\times m$ matrix $X^{(1)}$ is made from the lines $X$,  from the
second to the last one. Then denote
\[G^{(1)}=(M^{(1)}-z)^{-1}, \quad m^{(1)}=(M_{12},\dots M_{1n})\]
and use  (\ref{repr}) with these $G^{(1)}$ and $m^{(1)}$.

 To obtain the estimate for $\mathbf{E}_1\Big\{|\gamma_n-\mathbf{E}_{1}\{\gamma_n\}|^2\Big\}$
 we need (as in the proof of Proposition \ref{p:b_var}) to estimate
 $\mathbf{E}_{1}\{|A^\circ_1|^2\}/(\Im z \mathbf{E}_{1}\{A\})^2$ and
$\mathbf{E}_{1}\{|B^\circ_1|^2\}/( \mathbf{E}_{1}\{A\})^2$. Since $G^{(1)}$ does not depend on $\{X_{1i}\}_{i=1,.,m}$,
averaging with respect to $\{X_{1i}\}_{i=1,.,m}$, using the Jensen inequality and (\ref{BY}), we get
\begin{align*}
\mathbf{E}_{1}\{A\}&=\frac{1}{n}\mathrm{Tr\,} G^{(1)}M^{(1)},\quad \Im \mathbf{E}_{1}\{A\}=\frac{\Im z}{n}\mathrm{Tr\,}
G^{(1)}M^{(1)}G^{(1)*},\\
\mathbf{E}_{1}\{|A^\circ_1|^2\}&\le \frac{2+X_4}{n^2}\mathrm{Tr\,}
G^{(1)}M^{(1)}G^{(1)*}M^{(1)}\\&\le\frac{C}{n}\mathbf{E}_{1}^{1-\delta}\big\{n^{-1}\mathrm{Tr\,}
G^{(1)}M^{(1)}G^{(1)*}\big\}\mathbf{E}_{1}^{\delta}\big\{n^{-1}\mathrm{Tr\,}
G^{(1)}(M^{(1)})^{(1+\delta)/\delta}G^{(1)*}\big\}\\&\le\frac{C}{n|\Im
z|^{2\delta}}\mathbf{E}_{1}^{1-\delta}\big\{n^{-1}\mathrm{Tr\,} G^{(1)}M^{(1)}G^{(1)*}\big\}
\mathbf{E}_{1}^{\delta}\big\{n^{-1}\mathrm{Tr\,} (M^{(1)})^{(1+\delta)/\delta}\big\}.
\end{align*}
 But it is known (see \cite{Ba-Si:06} and references therein) that for any fixed $\delta>0$
\begin{equation}\label{BY}
\mathbf{E}\big\{n^{-1}\mathrm{Tr\,}(M^{(1)})^{(1+\delta)/\delta}\big\}\le
\big(2+\sqrt{c}\big)^{(1+\delta)/\delta}+o(1).
\end{equation}
Combining this bound with the above inequality and repeating the argument of Proposition \ref{p:b_var}, we obtain the
bound (\ref{b_v.2b}). The bound for $\mathbf{E}_{1}\{|B^\circ_1|^2\}/\mathbf{E}_{1}^2\{A\}$ can be obtained similarly.

\end{document}